\documentclass[sigconf,natbib=true]{acmart}
\AtBeginDocument{%
  \providecommand\BibTeX{{%
    \normalfont B\kern-0.5em{\scshape i\kern-0.25em b}\kern-0.8em\TeX}}}

\copyrightyear{2025}
\acmYear{2025}
\setcopyright{rightsretained}
\acmConference[WSDM '25]{Proceedings of the Eighteenth ACM International Conference on Web Search and Data Mining}{March 10--14, 2025}{Hannover, Germany}
   
\acmBooktitle{Proceedings of the Eighteenth ACM International Conference on Web Search and Data Mining (WSDM '25), March 10--14, 2025, Hannover, Germany}\acmDOI{10.1145/3701551.3703561} \acmISBN{979-8-4007-1329-3/25/03}

\makeatletter
\gdef\@copyrightpermission{
  \begin{minipage}{0.3\columnwidth}
   \href{https://creativecommons.org/licenses/by/4.0/}{\includegraphics[width=0.90\textwidth]{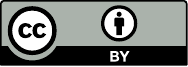}}
  \end{minipage}\hfill
  \begin{minipage}{0.7\columnwidth}
   \href{https://creativecommons.org/licenses/by/4.0/}{This work is licensed under a Creative Commons Attribution International 4.0 License.}
  \end{minipage}
  \vspace{5pt}
}
 
\usepackage{multirow}
\usepackage{makecell}
\usepackage{amsmath,amssymb,amsfonts}

\usepackage{array}   
\usepackage{multicol}
\usepackage{hyperref}
\usepackage{enumitem}

\usepackage{caption}

\begin{document}

\title{Spectrum-based Modality Representation Fusion Graph Convolutional Network for Multimodal Recommendation}

\author{Rongqing Kenneth Ong}
\affiliation{%
  \institution{Nanyang Technological University}
  \country{Singapore}
}
\email{rongqing001@e.ntu.edu.sg}
\author{Andy W.~H. Khong}
\affiliation{%
  \institution{Nanyang Technological University}
  \country{Singapore}
}
\email{andykhong@ntu.edu.sg}

\begin{abstract}
Incorporating multi-modal features as side information has recently become a trend in recommender systems. To elucidate user-item preferences, recent studies focus on fusing modalities via concatenation, element-wise sum, or attention mechanisms. Despite having notable success, existing approaches do not account for the modality-specific noise encapsulated within each modality. As a result, direct fusion of modalities will lead to the amplification of cross-modality noise. Moreover, the variation of noise that is unique within each modality results in noise alleviation and fusion being more challenging. In this work, we propose a new \underline{S}pectrum-based \underline{Mo}dality \underline{Re}presentation (SMORE) fusion graph recommender that aims to capture both uni-modal and fusion preferences while simultaneously suppressing modality noise. Specifically, SMORE projects the multi-modal features into the frequency domain and leverages the spectral space for fusion. To reduce dynamic contamination that is unique to each modality, we introduce a filter to attenuate and suppress the modality noise adaptively while capturing the universal modality patterns effectively. Furthermore, we explore the item latent structures by designing a new multi-modal graph learning module to capture associative semantic correlations and universal fusion patterns among similar items. Finally, we formulate a new modality-aware preference module, which infuses behavioral features and balances the uni- and multi-modal features for precise preference modeling. This empowers SMORE with the ability to infer both user modality-specific and fusion preferences more accurately. Experiments on three real-world datasets show the efficacy of our proposed model. The source code for this work has been made publicly available at~\url{https://github.com/kennethorq/SMORE}.
\end{abstract}



\maketitle

\section{Introduction}
In the rapidly expanding world of e-commerce, recommendation systems play a critical role in assisting users to identify products of interest. The standard user browsing experience encompasses exposure to various forms of multi-modal content~\cite{ma2024leveraging} intended to captivate and entice users. As user preferences are generally driven by a mixture of modality content, research into multi-modal recommender systems (MRSs) that leverage modalities to infer user interest has been gaining popularity. Studies have shown that the use of modalities outperforms general recommenders that rely solely on users' historical interaction~\cite{he2016vbpr, zhang2021mining, zhou2023bootstrap, zhou2023tale, kim2024monet}. 

The central focus of MRSs involves the integration of different modalities within the collaborative filtering (CF) framework. Earlier works such as VBPR~\cite{he2016vbpr} and DeepStyle~\cite{liu2017deepstyle} focus on fusing visual-specific modality by first projecting the modality features into a lower-dimensional space before combining it with the item identifier (ID) embeddings using concatenation and summation, respectively. Since user-item interactions can naturally be depicted as a bipartite graph~\cite{he2020lightgcn, dai2022towards, zhao2022joint, ong2023quad, chun2024reasonable, zheng2024missing}, graph neural networks (GNNs) have been adapted to capture the high-order connectivity of both multi-modal and behavioral interactions. To this end, LATTICE~\cite{zhang2021mining} constructs the latent structures associated with modalities by first constructing different views of the item graph. Thereafter, it performs modality fusion using an attention mechanism to weigh each modality. As an extension, FREEDOM~\cite{zhou2023tale} reduces redundancy in training the latent graphs by freezing them prior to training. More recently, to mitigate challenges associated with modality noise that may be introduced through pre-trained encoders, MGCN~\cite{yu2023multi} integrates behavioral features with modality features in an attempt to reduce such noise. It then employs average pooling to fuse each of the modality features.

\begin{figure}[t]
    \centering
    \includegraphics[scale=0.375]{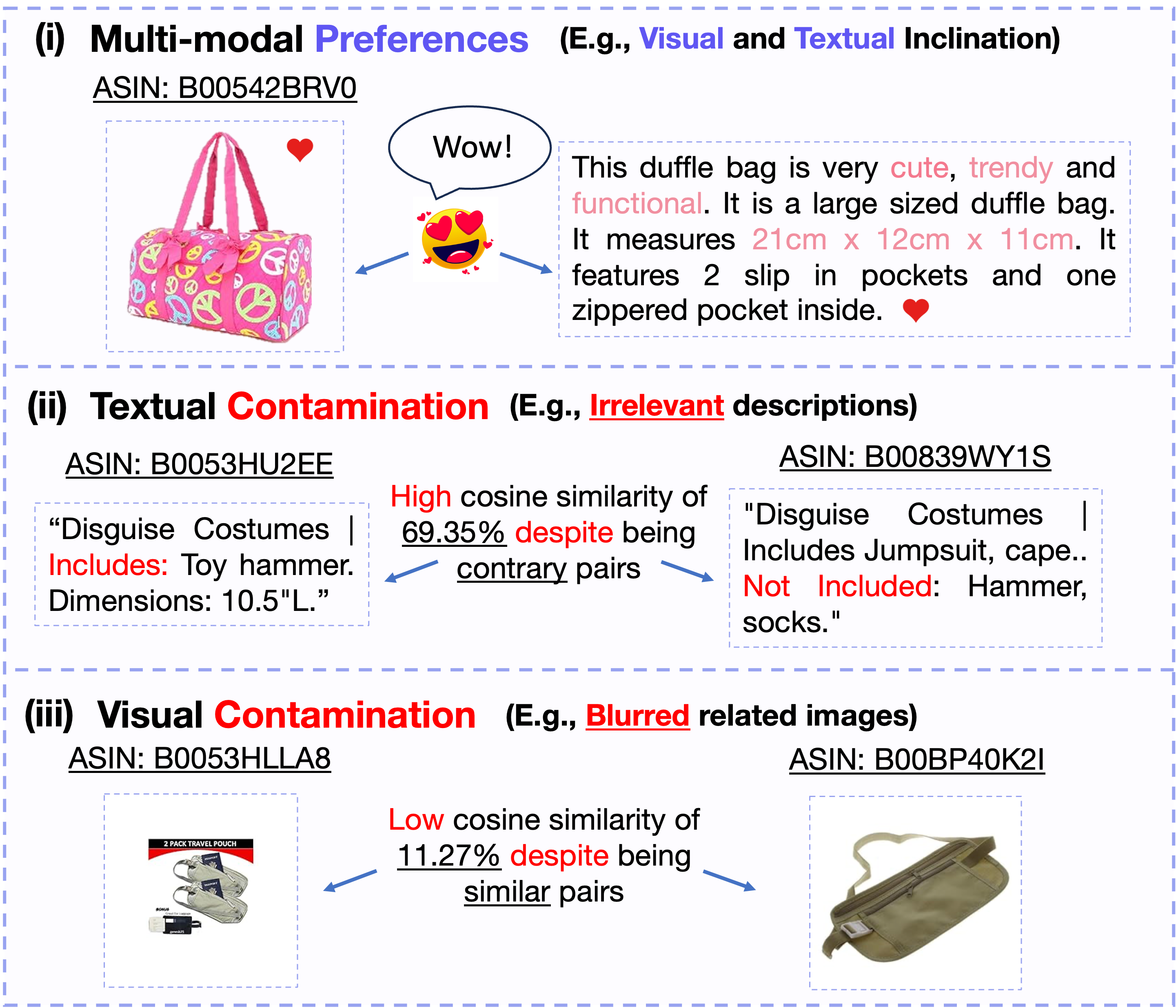}
    \caption{An illustrative example of user multi-modality preferences and issues related to modality-specific contamination. (i) User multi-modal preferences, (ii) irrelevant textual descriptions resulting in an unexpectedly high similarity score of 69.35\% even though item pairs are unrelated, and (iii) blurred images resulting in a low similarity score of 11.27\% even though pairs are related.}     
    \label{Case Study}
\end{figure}
While existing MRSs achieve notable success in incorporating multiple modalities, they suffer from a significant drawback\textemdash amplification of cross-modality noise during fusion~\cite{zhang2024multimodal,yang2023bicro}. Although it is essential to infer fusion preferences, existing works~\cite{he2016vbpr,liu2017deepstyle, zhang2022latent, zhang2021mining} combine modalities directly without taking into account the impact of modality-specific noise, which may detrimentally affect the quality of item representations learned. Consider an illustrative case study constructed from the Amazon Clothing~\cite{he2016ups} dataset. Fig.~\ref{Case Study}(i) highlights the importance of fusion preferences, where users make purchase decisions from an image and text description. However, capturing fusion preferences via modality fusion carries the risk of modality-specific contamination. For instance, consider a pair of items in (ii), which differ in terms of their functionality\textemdash the left item being a toy hammer, while the right being a jumpsuit. Due to the irrelevant description for the latter (which mentioned $Hammer$), these items yield a surprisingly high similarity score of 69.35\% despite them being uncorrelated. In terms of visual contamination, as shown in Fig.~\ref{Case Study}(iii), a similar pair of items (e.g., travel pouches) yields an unexpectedly low similarity score of 11.27\% when one of the images suffers from a blurring effect. As a result, the noise within each modality will be amplified further during fusion, thereby corrupting the item encoding process. Nonetheless, reducing noise contamination within each modality is highly challenging due to their unique and dynamic contamination characteristics~\cite{izadi2023image,anwar2019real}.
To reliably ascertain the multi-modal preferences of a user, a fusion module must be designed meticulously to mitigate the undesired effect of noise within each modality before fusion so as to capture universal patterns effectively while preserving essential uni-modal features.

In this work, we draw inspiration from the field of signal processing, which has shown to be effective for modality denoising~\cite{rao2021global, michelashvili2019speech}. By projecting signals to the frequency domain via the Fourier transform~\cite{heideman1984gauss}, a sparse frequency spectrum is generated. This unique characteristic facilitates the acquisition of discriminative spectrum~\cite{lao2024frequency}, which enables the distillation of critical modality features by means of effective attenuation. Furthermore, the frequency domain offers a comprehensive global perspective~\cite{rao2021global, xu2024sequence}, thereby enabling each spectral component to attend to all spatial domain features efficiently and effectively. Inspired by the discriminative spectral property and the global perspective by the frequency domain and to overcome the aforementioned drawbacks, we propose a new \underline{S}pectrum-based \underline{Mo}dality \underline{Re}presentation (SMORE) fusion graph recommender that aims to capture both uni-modal and fusion preferences while concurrently suppressing modality noise originating from raw features. In particular, SMORE comprises three key components: 1. Spectrum Modality Fusion, 2. Multi-modal Graph Learning, and 3. Modality-Aware Preference module. 

To capture the universal modality patterns holistically for inferring user fusion preferences, SMORE performs early fusion by projecting modality features into the frequency domain using the Fourier transform. Harnessing the global perspective inherent within the frequency domain, SMORE captures the cross-modality universal patterns effectively through an efficient point-wise aggregation. Given the discriminative spectrum features, a dynamic filter is then formulated to attenuate and suppress irrelevant (noise) signals adaptively during the fusion process, ensuring that only essential sequence and spatial features are transmitted and fused.

Furthermore, we exploit the correlations between collaborative and latent structures by designing a new multi-modal graph learning module to encode high-order collaborative and relational signals from two distinct perspectives: user-item and item-item modality views, respectively. A new modality-aware preference module is also proposed to capture users' uni- and multi-modal preferences comprehensively. By injecting behavioral signals into the uni-modal and fusion features, SMORE effectively balances and achieves a more concise modeling of preferences between the uni-modal and fusion content. Experiments conducted on three real-world datasets validate the efficacy of our proposed model.

The contributions of our work are threefold:
\begin{itemize}
    \item We propose a new spectrum-based modality fusion scheme to fuse modalities associated with different semantics effectively while suppressing the modal-specific noise from its raw content;
    \item We design a multi-modal graph learning module comprising modal-specific and -fusion views to capture high-order collaborative and semantically correlated signals;
    \item We formulate a new modality-aware preference module to capture the users' diverse uni- and multi-modal preferences explicitly, reflecting real-world scenarios.
\end{itemize}

\begin{figure*}[t]
    \centering
    \includegraphics[width=1\textwidth, height=0.5\textheight, keepaspectratio]{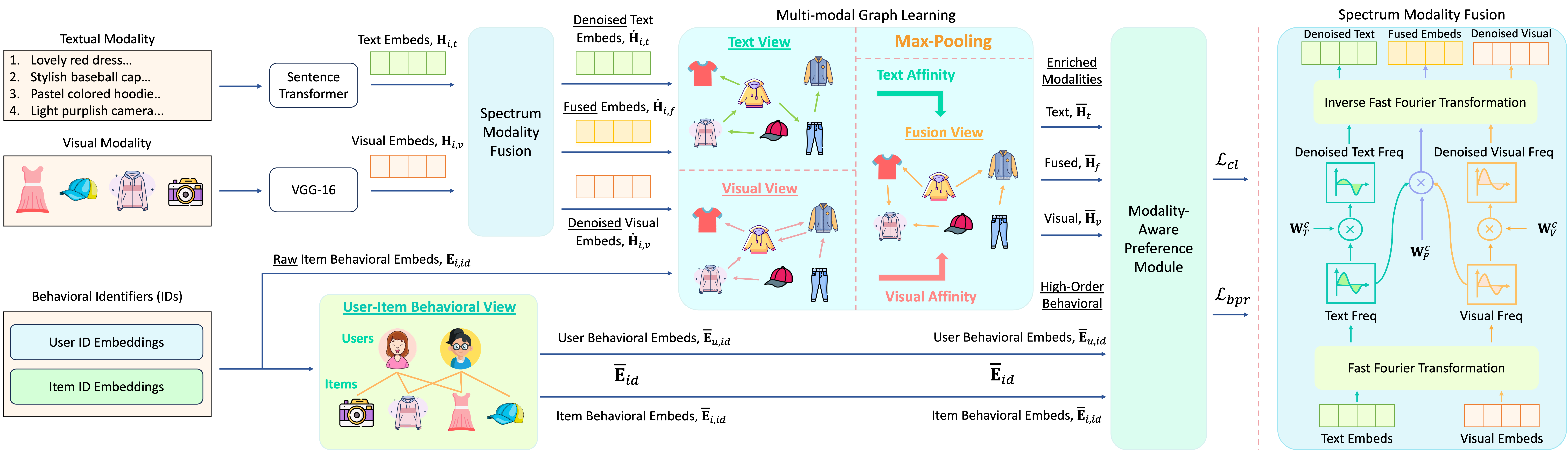}
    \caption{An illustrative overview of the proposed architecture, comprising three key components: (i) spectrum modality fusion, (ii) multi-modal graph learning, and (iii) modality-aware preference module.}
    \label{SMORE Architecture}
\end{figure*}

\section{Related Works}
\subsection{Multi-modal Recommendation}
Integrating semantically rich multi-modality features into recommender systems has recently emerged as a predominant way to enhance the accuracy of recommendation systems. One of the conventional approaches to extract multi-modal features is through the use of pre-trained neural networks (e.g., Sentence Transformer~\cite{reimers2019sentence}, VGG-16~\cite{simonyan2014very}). For instance, VBPR~\cite{he2016vbpr} employs a pre-trained convolutional neural network (CNN) to extract deep visual features corresponding to the items. Consequently, it performs modality fusion by concatenating the ID and visual embeddings to model user modality-specific preferences. VECF~\cite{chen2019personalized}, on the other hand, leverages VGG-16 to analyze users’ complex preferences for image patches by applying pre-segmentation~\cite{ronneberger2015u, chen2024adaptive} and an attention model to capture key regions within images.

As user interactions naturally occur in the form of structured data, recent works utilize GNNs to capture such structural information in MRSs. Leveraging multi-modal features (e.g., visual, text, acoustic), DualGNN~\cite{wang2021dualgnn} integrates a user co-occurrence graph and a preference learning module to model different granularities of user preferences. By introducing behavioral information into modality features, MGCN~\cite{yu2023multi} incorporates an attention layer to capture the importance of different modalities. BM3~\cite{zhou2023bootstrap}, on the other hand, adopts a new self-supervised learning approach that aims to reduce computational resource demands and rectify incorrect supervision signals arising from negative sampling. While proven effective, these models treat modalities independently and do not account for the fusion preferences that a user may have. 

\subsection{Modality Fusion Graph-based Learning}
Apart from creating modalities of different views via graph neural networks, several graph-based approaches have attempted to fuse the multi-modalities directly. In essence, fusion-based approaches can be categorized into three main stages: early, intermediate, and late fusion~\cite{zhou2023comprehensive}. For instance, LATTICE~\cite{zhang2021mining} performs early fusion by first creating similarity graphs of different modalities to preserve essential item-item connections. Thereafter, a weighted-sum fusion is applied using an importance score before the use of graph convolution to obtain the final fused representation. As an extension, FREEDOM~\cite{zhou2023tale} reveals the redundancy in learning item-item similarity graphs and freezes the modality graphs during fusion for better representation learning. In terms of intermediate fusion, MMGCL~\cite{yi2022multi} augments multi-modal graphs with edge dropout and masking, followed by concatenation and message propagation through a graph encoder. Instead of direct concatenation, DRAGON~\cite{zhou2023enhancing} adopts the late fusion paradigm and introduces attentive concatenation to discern user preferences across varying modalities. While these fusion approaches have achieved some success, noise embedded within modalities has not been effectively suppressed, resulting in noise amplification and degradation of model performance during cross-modality fusion. In this work, we propose to fuse and denoise modalities from the frequency spectrum perspective\textemdash an effective yet under-explored area in MRSs.

\section{Task Formulation}
We model users' implicit interaction by first defining~$\mathcal{U} = \{ u_j \mid 1 \leq j \leq M \}$ and $\mathcal{I} = \{ i_k \mid 1 \leq k \leq N \}$ as the user and item sets,  where $M$ and $N$ denote the number of users and items, respectively. Henceforth, we can then define an interaction matrix $\mathbf{Y}\in\mathbb{R}^{M\times N}$, where element $y_{ui}=1$ signifies an observed user-item interaction, while $y_{ui}=0$ indicates otherwise. For each user $u$ and item $i$, the input ID embedding matrix is represented as $E_{id}\in\mathbb{R}^{d\times\left( |U| + |I|\right)}$, where $d$ is the embedding dimension. We further denote $e_{i}^{m}\in\mathbb{R}^{d_{m}}$ as the modality features of each item $i$, where $d_{m}$ is the dimension of the modalities, $m\in\mathcal{M}$ is the modality, and $\mathcal{M}$ is the set of modalities. 
The primary focus of this work is on visual and textual modalities, where $\mathcal{M}=\{v, t\}$ such that $v$ and $t$ correspond, respectively, to visual and textual modality. While two modalities are described in this work, the proposed approach can easily be extended to multiple modalities. Finally, given the interaction data and the multi-modal features of each item, our goal is to predict user preferences accurately by estimating the likelihood $\widehat{y}_{ui}$ of interaction between a user $u$ and an item $i$.

\section{Methodology}
The key components of SMORE encompass three core aspects: i) Spectrum Modality Fusion, ii) Multi-modal Graph Learning, and iii) Modality-aware Preference Module. Fig.~\ref{SMORE Architecture} illustrates an overview of the proposed architecture.

\subsection{Spectrum Modality Fusion}
Modality fusion often confers advantages since the fused embeddings elucidate complementary and universal characteristics of different modalities. As opposed to existing works and drawing inspiration from the field of signal processing, SMORE exploits the frequency domain for dual purposes: modality fusion and denoising. With this objective, the raw multi-modal features $\mathbf{E}_{i,m}$ are first projected into a shared latent space using the multi-layer perceptron (MLP), i.e.,
\small
\begin{equation}
\mathbf{H}_{i,m} = \mathbf{W}_{1,m} \mathbf{E}_{i,m} + \mathbf{b}_{1,m},
\end{equation}
\normalsize
where $\mathbf{W}_{1,m}\in\mathbb{R}^{d\times d_{m}}$ and $\mathbf{b}_{1,m}\in\mathbb{R}^{d}$ denote the projection matrix and bias vector of the MLP for each modality $m$, respectively. Thereafter, to convert the projected multi-modal features into the frequency domain for fusion and denoising, we utilize the discrete Fourier transform (DFT)~\cite{heideman1984gauss} for each modality such that 
\small
\begin{equation}
\widetilde{\mathbf{H}}_{i,m} = \mathcal{F}_m\left(\mathbf{H}_{i,m}\right) \in \mathbb{C}^{n \times d},
\end{equation}
\normalsize
where  
\small
\begin{equation}
\mathcal{F}_m : \widetilde{h}_k = \sum_{j=0}^{n-1} h_j \exp \left( -\frac{2\pi \jmath}{n} jk \right), \quad 0 \leq k \leq n-1
\end{equation}
\normalsize
denotes the one-dimensional DFT function along the sequence and spatial dimensions of the textual and image modality, respectively, $\jmath=\sqrt{-1}$, and $\widetilde{h}_k$ denotes the modality spectrum features at frequency $2\pi k / {n}$, with $k$ being the frequency-bin index. While the DFT has been widely applied for frequency conversion, quadratic complexity is incurred due to the computation of $N$ components. Instead, we employ the fast Fourier transform (FFT), which decomposes the DFT matrix into a series of sparse matrix products~\cite{van1992computational}, thereby reducing the complexity to a logarithmic scale.

With the spectral features of each modality, denoising is performed. Since dynamic noise may be present across different modalities and leveraging the discriminative spectrum generated from the FFT~\cite{lao2024frequency},  we introduce a modality-specific dynamic filter
\small
\begin{equation}
\mathbf{\widehat{H}}_{i,m} = \delta_{m} \left( \widetilde{\mathbf{H}}_{i,m} \right) = \mathbf{W}_{2,m}^{c} \odot \mathbf{\widetilde{H}}_{i,m},
\end{equation}
\normalsize
where $\delta_{m}$ is defined as the modal-specific transfer function of the filter, $\odot$ denotes the point-wise product, and $\mathbf{W}_{2,m}^{c}\in\mathbb{C}^{n \times d}$ denotes the trainable complex weight of the filter. The adaptive filter functions as a frequency selector that seeks to suppress noise-related irrelevant information. We can formulate the fusion process similarly to acquire the cross-modality fusion spectrum such that
\small
\begin{equation}
\mathbf{\widehat{H}}_{i,f} = \delta_{f} \left( \mathop{\mathbf{\Pi}}\limits_{m \in \mathcal{M}} \mathbf{\widetilde{H}}_{i,m} \right),
\end{equation}
\normalsize
where, as opposed to matrix multiplication, $\mathbf{\Pi}$ is defined as the point-wise product operator, and $\delta_{f}$ is defined as the transfer function of the dynamic fusion filter. It is important to note that in the frequency domain, the point-wise product operation is equivalent to the circular convolution operation in the spatial domain. As a result, the rich correlations between the sequence and spatial modality (e.g., text and image) are captured, 
while minimizing noise contamination during fusion. More importantly, in contrast to advanced fusion methods (e.g., co-attention~\cite{wei2023multi}), which require a quadratic time complexity, fusing in the frequency domain allows SMORE to achieve logarithmic runtime due to the efficient FFT and point-wise aggregation. In this aspect, we can achieve both efficiency and effectiveness in fusing modalities and denoising.

Thereafter, the spectrum of the uni-modal and fused modality features are projected back into the original feature space using the inverse discrete Fourier transform (IDFT)
\small
\begin{equation}
\mathcal{F}^{~-1}_{m} : h_j = \frac{1}{n} \sum_{k=0}^{n-1} \tilde{h}_k \exp \left( \frac{2\pi \jmath}{n} jk \right), \quad 0 \leq j \leq n-1,
\end{equation}
\begin{equation}
\dot{\mathbf{H}}_{i,m} = \mathcal{F}^{~-1}_{m}(\mathbf{\widehat{H}}_{i,m}) \in \mathbb{R}^{n \times d}, \quad \dot{\mathbf{H}}_{i,f} = \mathcal{F}^{~-1}_{m}(\mathbf{\widehat{H}}_{i,f}) \in \mathbb{R}^{n \times d}.
\end{equation}
\normalsize
As will be illustrated in Section~\ref{sec:fusion}, the above empowers SMORE to execute modality fusion and denoising effectively, extracting only essential uni-modal and fused features through filtering in the frequency domain. 

\subsection{Multi-modal Graph Learning}
\subsubsection{\textbf{Item-Item Modal-Specific and Fusion Views.}}
Having acquired the denoised and fused modality representations, the semantically correlated modality features can be distilled through graph convolutional operations. As highlighted in~\cite{zhang2021mining, zhang2022latent, zhou2023tale}, the efficacy of a multi-modal recommender can be influenced substantially by both collaborative and semantically associated signals. These works construct individual graphs for each modality and aggregate them via learnable weights before performing message propagation on the fusion graph. By aggregating and disregarding uni-modality graph, distinct modality preferences are obscured. 

In contrast, we emphasize the importance of capturing both uni-modal and fusion preferences by proposing a new multi-modal graph learning module that distinctively constructs modal-specific and fusion graphs. We first establish item-item affinities by computing the similarity of each raw modality features. Henceforth, we attain modality similarity matrix $\mathcal{S}_m$ such that the similarity between (item) row $a$ and (item) column $b$ entry of $\mathbf{E}_{i,m}$ is given by
\small
\begin{equation}
s_{a,b}^m = \frac{\left( e_a^m \right)^{\text{T}} e_b^m}{\| e_a^m \| \| e_b^m \|}.
\end{equation}
\normalsize
To ensure that the uni-modal vital features are captured, we perform graph sparsification~\cite{chen2009fast} by retaining $K$ edges with the highest similarity scores such that
\small
\begin{equation}
\label{knn_sparsification}
\dot{s}_{a,b}^m =
\begin{cases} 
s_{a,b}^m, & s_{a,b}^m \in \text{top-}K_m(\{s_{a,c}^m, \, c \in \mathcal{I}\}); \\
0, & \text{otherwise},
\end{cases}
\end{equation}
\normalsize
where $\dot{s}_{a,b}^m$ denotes the degree of similarity (edge weights) between items $a$ and $b$ in modality $m$. To mitigate the gradient vanishing/exploding problem~\cite{kipf2016semi}, we then normalize the similarity matrix
\small
\begin{equation}
\ddot{\mathbf{\mathcal{S}}}_m = \mathbf{D}_m^{-1/2} \dot{\mathbf{\mathcal{S}}}_m \mathbf{D}_m^{-1/2},
\end{equation}
\normalsize
where $\mathbf{D}_m^{-1/2}$ is defined as the degree matrix of $\dot{\mathbf{\mathcal{S}}}_m$. 

Having constructed the modality-specific graph, we adopt the max-pooling strategy to retain the highest complementary strength between different $m$ modality graphs, thereby preserving prominent cross-modality features. The fusion affinity matrix can be defined as the $\max$ edge weights between items $a$ and $b$, i.e.,
\small
\begin{equation}
\ddot{\mathbf{\mathcal{S}}}_{a,b}^{f} = \max_{m,m' \in \mathcal{M}} \left( \ddot{\mathbf{\mathcal{S}}}_{a}^{m}, \ddot{\mathbf{\mathcal{S}}}_{b}^{m'} \right), \quad m \neq m'.
\end{equation}
\normalsize
Prior to unimodal and fusion feature propagation, we extract preference-related modality features based on behavioral guidance
\small
\begin{align}
\ddot{\mathbf{H}}_{i,m} = &f_{gate}^{m}\left( \mathbf{E}_{i,id}, \dot{\mathbf{H}}_{i,m} \right) = \mathbf{E}_{i,id} \odot \sigma \left( \mathbf{W}_{3,m} \dot{\mathbf{H}}_{i,m} + \mathbf{b}_{3,m} \right), \\
\ddot{\mathbf{H}}_{i,f} = &f_{gate}^{c}\left( \mathbf{E}_{i,id}, \dot{\mathbf{H}}_{i,f} \right) = \mathbf{E}_{i,id} \odot \sigma \left( \mathbf{W}_{4,f} \dot{\mathbf{H}}_{i,f} + \mathbf{b}_{4,f} \right),
\end{align}
\normalsize
where $\mathbf{W}_{(.)}\in\mathbb{R}^{d\times d}$ and $\mathbf{b}_{(.)}\in\mathbb{R}^{d}$ are the trainable parameters 
and $\sigma$ is the non-linearity sigmoid gate function. 

Inspired by the simplicity and efficacy of LightGCN~\cite{he2020lightgcn}, the item uni-modal $\ddot{\mathbf{H}}_{i,m}$ and fusion features $\ddot{\mathbf{H}}_{i,f}$ are propagated through a shallow light graph convolutional layer with the propagation rule being 
\small
\begin{equation}
\widebar{\mathbf{H}}_{i,m} = \ddot{\mathbf{\mathcal{S}}}_m \ddot{\mathbf{H}}_{i,m}, \qquad \widebar{\mathbf{H}}_{i,f} = \ddot{\mathbf{\mathcal{S}}}_f \ddot{\mathbf{H}}_{i,f}.
\end{equation}
\normalsize
Likewise, we can compute the user modality features through a weighted-sum aggregation layer defined by 
\small
\begin{equation}
\widebar{h}_{u,m} = \sum_{i \in \mathcal{N}_u} \frac{1}{\sqrt{|\mathcal{N}_u| |\mathcal{N}_i|}} \widebar{h}_{i,m}, \qquad \widebar{h}_{u,f} = \sum_{i \in \mathcal{N}_u} \frac{1}{\sqrt{|\mathcal{N}_u| |\mathcal{N}_i|}} \widebar{h}_{i,m},
\end{equation}
\normalsize
where $\widebar{h}_{u,m}$ and $\widebar{h}_{u,f}$ are the user uni and fusion modality features, respectively, and $1 / \sqrt{\scalebox{0.9}{$|\mathcal{N}_u| |\mathcal{N}_i|$}}$ is the symmetric normalization term to avoid overscaling. It is useful to note that employing a shallow layer in the item-item view is sufficient for capturing relevant semantic associative signals since stacking multiple layers may induce undesirable high-order latent noise~\cite{yu2023multi}. By concatenating $\widebar{\mathbf{H}}_{u,m}$ with $\widebar{\mathbf{H}}_{i,m}$, and $\widebar{\mathbf{H}}_{u,f}$ with $\widebar{\mathbf{H}}_{i,f}$, we can obtain the enriched uni-modal and fusion features for both the users and items, denoted as $\widebar{\mathbf{H}}_{m}$ and $\widebar{\mathbf{H}}_{f}\in \mathbb{R}^{d \times (|\mathcal{U}|+|\mathcal{I}|)}$, respectively.

\subsubsection{\textbf{User-Item Behavioral View.}}
The focus of this view is on encoding the high-order collaborative signals from users' historical interactions. It has been verified that the collaborative signals are highly influential in delineating users' behavioral patterns~\cite{mao2021ultragcn,he2020lightgcn, zhou2023selfcf, he2023simplifying}. On this basis, we recursively propagate long-range collaborative signals in the interaction graph resulting in the behavioral embedding of the users and items given by
\small
\begin{equation}
\mathbf{E}_{id}^{(l)} = (\mathbf{D}^{-1/2} \mathbf{A} \mathbf{D}^{-1/2}) \: \mathbf{E}_{id}^{(l-1)}, \quad \mathbf{A} = \begin{bmatrix} 0 & \mathbf{Y} \\ \mathbf{Y}^\top & 0 \end{bmatrix}.
\end{equation}
\normalsize
Here, $\mathbf{E}_{id}^{(l-1)}$ denotes the ID embeddings at the previous layer, and $\mathbf{D}^{-1/2}$ is the diagonal degree matrix corresponding to the adjacency matrix $\mathbf{A}$. To obtain the overall high-order behavioral features of users and items, we aggregate the hidden layers by applying the mean function giving
\small
\begin{equation}
\widebar{\mathbf{E}}_{id} = \frac{1}{L+1} \sum_{i=0}^{L} \mathbf{E}_{id}^{(l)}, \qquad \widebar{\mathbf{E}}_{id} \in \mathbb{R}^{d \times (|\mathcal{U}|+|\mathcal{I}|)}.
\end{equation}
\normalsize

\subsection{Modality-Aware Preference Module}
With the high-order behavioral and enriched (uni-modal and fused) features $\widebar{\mathbf{E}}_{id}$, $\widebar{\mathbf{H}}_{m}$ and $\widebar{\mathbf{H}}_{f}$, modality preferences are distilled for each user. In line with the diversity observed in real-world scenarios, a user may be inclined toward a single modality, while some may exhibit a mixture of fusion preferences. As such, we utilize complementary signals encapsulated in the fusion embeddings to weigh the uni-modal features, effectively striking a balance between the uni-modal and fusion preferences. To this end, we define
\small
\begin{equation}
\alpha_m = \text{softmax}(\mathbf{p}_m^\top \tanh(\mathbf{W}_{5,m} \widebar{\mathbf{H}}_f + \mathbf{b}_{5,m}))
\end{equation}
\normalsize
as the modal-specific attention weights, where $p_{(\cdot)} \in \mathbb{R}^d$ is the attention vector. These weights are subsequently used to weigh the semantically associated uni-modal features to obtain the final aggregated uni-modal features given by
\small
\begin{equation}
\mathbf{H}^{*}_{m} = \sum_{m \in M} \alpha_m \widebar{\mathbf{H}}_m.
\end{equation}
\normalsize

We next extract modality preferences derived from user collaborative information by feeding the high-order behavioral signal through a uni-modal and fusion gate function. This results in explicit uni-modal and fusion preferences given, respectively, by
\small
\begin{eqnarray}
\begin{aligned}
        &\mathbf{Q}_m = \psi_{pref}^{m} \left( \widebar{\mathbf{E}}_{id} \right) = \sigma(\mathbf{W}_{6,m} \widebar{\mathbf{E}}_{id} + \mathbf{b}_{6,m}),\\
        &\mathbf{Q}_f = \psi_{pref}^{f} \left( \widebar{\mathbf{E}}_{id} \right) = \sigma(\mathbf{W}_{7,f} \widebar{\mathbf{E}}_{id} + \mathbf{b}_{7,f}),
\end{aligned}
\end{eqnarray}
\normalsize
where $\sigma$ is the non-linearity function. The overall multi-modal side features (distilled from the explicit uni-modal and fusion preferences) are then derived as
\small
\begin{equation}
\mathbf{H}_{s} = \frac{1}{|M|} \left( \sum_{m \in M} \mathbf{H}^{*}_m \odot \mathbf{Q}_m \right) +  \left( \mathbf{H}_f \odot \mathbf{Q}_{f} \right).
\end{equation}
\normalsize

To maximize mutual information across high-order behavioral and modality-side information, we then incorporate an InfoNCE contrastive task~\cite{oord2018representation} with
\small
\begin{equation}
\mathcal{L}_{cl}^{u} = \sum_{u \in \mathcal{U}} -\log \frac{\exp(\widebar{e}_{u,id} \cdot \widebar{h}_{u,s} / \tau)}{\sum_{v \in \mathcal{U}} \exp(\widebar{e}_{v,id} \cdot \widebar{h}_{v,s} / \tau)}
\label{cl_eq}
\end{equation}
\normalsize
being the user contrastive loss and $\tau$ being the hyperparameter temperature that regulates the degree of smoothness in the distribution. This task ensures the preservation of essential features distilled from behavioral and modality views. Similarly, we can obtain the item contrastive loss  $\mathcal{L}_{cl}^{i}$ by substituting users with items as defined in Eq~(\ref{cl_eq}). Thereafter, the overall contrastive loss is governed by $\mathcal{L}_{cl}=\mathcal{L}_{cl}^{u} + \mathcal{L}_{cl}^{i}$.

\subsection{Prediction and Optimization}
By capitalizing on the refined uni-modal and complementary fused features, we acquire the final representations of the user and item
\small
\begin{equation}
e_u^{*} = \widebar{e}_{u,id} + h_{u,s}, \quad e_i^{*} = \widebar{e}_{i,id} + h_{i,s}.
\end{equation}
\normalsize
The estimated likelihood is then computed as $\widehat{y}(u,i) =e_u^{*\top} e_i$. For model optimization, we employ the BPR loss~\cite{rendle2012bpr} to reconstruct the historical data, which prioritizes higher scores for observed items, i.e.,
\small
\begin{equation}
\mathcal{L}_{bpr}=\sum_{(u,i,j)\in O} -\mathrm{ln} \space \sigma\Big(\widehat{y}_{ui}-\widehat{y}_{uj}\Big).
\end{equation}
\normalsize
Here, $O = \{(u,i,j)|(u,i)\in O^{+},(u,j)\in O^{-}\}$ represents the set of interactions, comprising observed $O^{+}$ and unobserved $ O^{-}$ interactions, and $\sigma$ denotes the sigmoid function. We then perform joint optimization in conjunction with the contrastive loss such that the overall loss function is given by
\small
\begin{equation}
\mathcal{L} = \mathcal{L}_{bpr} + \lambda_{1} \mathcal{L}_{cl} + \lambda_{2} ||\Theta||_{2}^{2},
\end{equation}
\normalsize
where $\lambda_{1}$ and $\lambda_{2}$ regulates the influence of the contrastive task and the L2 regularization term, respectively.
\begin{table}[t]
\caption{Dataset Statistics}
\label{Raw_Dataset}
\centering
\begin{tabular}{lrrrr}
\toprule
Dataset & \#User & \#Item & \#Interaction & Density \\
\midrule
Baby & 19,445 & 7,050 & 160,792 & 0.117\% \\
Sports & 35,598 & 18,357 & 296,337 & 0.045\% \\
Clothing & 39,387 & 23,033 & 278,677 & 0.031\% \\
\bottomrule
\end{tabular}
\end{table}
{
\begin{table*}[ht]
\centering
\caption{Performance comparison of different recommendation models. To ascertain the stability of the results, experiments were conducted across 5 different seeds, and the improvements are statistically significant with $p$ < 0.01 in a paired t-test setting.}
\label{tab: Overall Performance Comparison}
\begin{tabular}{l l l l l l l l l l l l l l l l l  l  l l l l }
\toprule
\multirow{2}{*}{Datasets} & \multirow{2}{*}{Metrics} & \multicolumn{4}{c}{General Recommenders} & \multicolumn{14}{c}{Multi-modal Recommenders} \\ 
\cmidrule(lr){3-6} \cmidrule(lr){7-22}
& & \multicolumn{2}{c}{BPR} & \multicolumn{2}{c}{LightGCN} & \multicolumn{2}{c}{VBPR} & \multicolumn{2}{c}{MMGCN} & \multicolumn{2}{c}{GRCN} & \multicolumn{2}{c}{SLMRec} & \multicolumn{2}{c}{BM3} & \multicolumn{2}{c}{MGCN} & \multicolumn{2}{c}{FREEDOM} & \multicolumn{2}{c}{SMORE} \\ \midrule
\multirow{4}{*}{Baby} & Recall@10 & \multicolumn{2}{c}{0.0382} & \multicolumn{2}{c}{0.0453} & \multicolumn{2}{c}{0.0425} & \multicolumn{2}{c}{0.0424} & \multicolumn{2}{c}{0.0534} & \multicolumn{2}{c}{0.0545} & \multicolumn{2}{c}{0.0548} & \multicolumn{2}{c}{0.0616} & \multicolumn{2}{c}{\underline{0.0626}} & \multicolumn{2}{c}{\textbf{0.0680*}} \\ %
& Recall@20 & \multicolumn{2}{c}{0.0595} & \multicolumn{2}{c}{0.0728} & \multicolumn{2}{c}{0.0663} & \multicolumn{2}{c}{0.0668} & \multicolumn{2}{c}{0.0831} & \multicolumn{2}{c}{0.0837} & \multicolumn{2}{c}{0.0876} & \multicolumn{2}{c}{0.0943} & \multicolumn{2}{c}{\underline{0.0986}} & \multicolumn{2}{c}{\textbf{0.1035*}} \\ %
& NDCG@10 & \multicolumn{2}{c}{0.0207} & \multicolumn{2}{c}{0.0246} & \multicolumn{2}{c}{0.0223} & \multicolumn{2}{c}{0.0223} & \multicolumn{2}{c}{0.0288} & \multicolumn{2}{c}{0.0296} & \multicolumn{2}{c}{0.0297} & \multicolumn{2}{c}{\underline{0.0330}} & \multicolumn{2}{c}{0.0327} & \multicolumn{2}{c}{\textbf{0.0365*}} \\ %
& NDCG@20 & \multicolumn{2}{c}{0.0263} & \multicolumn{2}{c}{0.0317} & \multicolumn{2}{c}{0.0284} & \multicolumn{2}{c}{0.0286} & \multicolumn{2}{c}{0.0365} & \multicolumn{2}{c}{0.0371} & \multicolumn{2}{c}{0.0381} & \multicolumn{2}{c}{0.0414} & \multicolumn{2}{c}{\underline{0.0420}} & \multicolumn{2}{c}{\textbf{0.0457*}} \\ \midrule%
\multirow{4}{*}{Sports} & Recall@10 & \multicolumn{2}{c}{0.0417} & \multicolumn{2}{c}{0.0542} & \multicolumn{2}{c}{0.0561} & \multicolumn{2}{c}{0.0386} & \multicolumn{2}{c}{0.0607} & \multicolumn{2}{c}{0.0676} & \multicolumn{2}{c}{0.0613} & \multicolumn{2}{c}{\underline{0.0736}} & \multicolumn{2}{c}{0.0724} & \multicolumn{2}{c}{\textbf{0.0762*}} \\ %
& Recall@20 & \multicolumn{2}{c}{0.0633} & \multicolumn{2}{c}{0.0837} & \multicolumn{2}{c}{0.0857} & \multicolumn{2}{c}{0.0627} & \multicolumn{2}{c}{0.0922} & \multicolumn{2}{c}{0.1017} & \multicolumn{2}{c}{0.0940} & \multicolumn{2}{c}{\underline{0.1105}} & \multicolumn{2}{c}{0.1089} & \multicolumn{2}{c}{\textbf{0.1142*}} \\ %
& NDCG@10 & \multicolumn{2}{c}{0.0232} & \multicolumn{2}{c}{0.0300} & \multicolumn{2}{c}{0.0307} & \multicolumn{2}{c}{0.0204} & \multicolumn{2}{c}{0.0325} & \multicolumn{2}{c}{0.0374} & \multicolumn{2}{c}{0.0339} & \multicolumn{2}{c}{\underline{0.0403}} & \multicolumn{2}{c}{0.0390} & \multicolumn{2}{c}{\textbf{0.0408*}} \\ %
& NDCG@20 & \multicolumn{2}{c}{0.0288} & \multicolumn{2}{c}{0.0376} & \multicolumn{2}{c}{0.0384} & \multicolumn{2}{c}{0.0266} & \multicolumn{2}{c}{0.0406} & \multicolumn{2}{c}{0.0462} & \multicolumn{2}{c}{0.0424} & \multicolumn{2}{c}{\underline{0.0498}} & \multicolumn{2}{c}{0.0484} & \multicolumn{2}{c}{\textbf{0.0506*}} \\ \midrule%
\multirow{4}{*}{Clothing} & Recall@10 & \multicolumn{2}{c}{0.0200} & \multicolumn{2}{c}{0.0338} & \multicolumn{2}{c}{0.0281} & \multicolumn{2}{c}{0.0224} & \multicolumn{2}{c}{0.0428} & \multicolumn{2}{c}{0.0461} & \multicolumn{2}{c}{0.0418} & \multicolumn{2}{c}{\underline{0.0649}} & \multicolumn{2}{c}{0.0635} & \multicolumn{2}{c}{\textbf{0.0659*}} \\%
& Recall@20 & \multicolumn{2}{c}{0.0295} & \multicolumn{2}{c}{0.0517} & \multicolumn{2}{c}{0.0410} & \multicolumn{2}{c}{0.0362} & \multicolumn{2}{c}{0.0663} & \multicolumn{2}{c}{0.0696} & \multicolumn{2}{c}{0.0636} & \multicolumn{2}{c}{\underline{0.0971}} & \multicolumn{2}{c}{0.0938} & \multicolumn{2}{c}{\textbf{0.0987*}} \\%
& NDCG@10 & \multicolumn{2}{c}{0.0111} & \multicolumn{2}{c}{0.0185} & \multicolumn{2}{c}{0.0157} & \multicolumn{2}{c}{0.0118} & \multicolumn{2}{c}{0.0227} & \multicolumn{2}{c}{0.0249} & \multicolumn{2}{c}{0.0225} & \multicolumn{2}{c}{\underline{0.0356}} & \multicolumn{2}{c}{0.0340} & \multicolumn{2}{c}{\textbf{0.0360*}} \\%
& NDCG@20 & \multicolumn{2}{c}{0.0135} & \multicolumn{2}{c}{0.0230} & \multicolumn{2}{c}{0.0190} & \multicolumn{2}{c}{0.0153} & \multicolumn{2}{c}{0.0287} & \multicolumn{2}{c}{0.0308} & \multicolumn{2}{c}{0.0281} & \multicolumn{2}{c}{\underline{0.0438}} & \multicolumn{2}{c}{0.0417} & \multicolumn{2}{c}{\textbf{0.0443*}} \\ \bottomrule
\end{tabular}
\end{table*}
}

\section{Experiments}
We conducted an extensive set of experiments designed to address the following research questions:
\begin{itemize}
    \item \textbf{RQ1:} How effective is the proposed SMORE architecture compared with state-of-the-art general and multi-modal models?
    \item \textbf{RQ2:} How do the key components and different modalities within SMORE contribute to its overall performance?
    \item \textbf{RQ3:} How do hyperparameter perturbations impact the overall efficacy of the proposed model?
    \item \textbf{RQ4:} Does spectrum-based fusion truly enhance denoising capability and capture valuable content?
\end{itemize}
\subsection{Experiment Configurations}
\subsubsection{Datasets.}
In accordance with preceding works~\cite{zhang2021mining, zhou2023bootstrap}, we perform experiments using three categories of the real-world Amazon Review datasets\footnote{Datasets are publicly available at~\url{http://jmcauley.ucsd.edu/data/amazon/links.html}.}, presented by McAuley et al.~\cite{he2016vbpr}: (i) $Baby$, (ii) $Sports\,and\,Outdoors$, and (iii) $Clothing,\,Shoes\,and\,Jewelry$. For ease of reference, we label them as $Baby$, $Sports$, and $Clothing$, respectively. Offering both visual and textual insights into items, the Amazon dataset exhibits variability in the number of items per category. For pre-processing, we filter the raw data from each dataset using the 5-core setting on both users and items. The data has been summarized in Table~\ref{Raw_Dataset}. For the visual modality, we adopted the 4,096-dimensional features acquired from VGG16~\cite{simonyan2014very}. For the textual modality, we employed sentence-transformers~\cite{reimers2019sentence} to obtain a 384-dimensional text embedding from the concatenated brand, title, description, and category of each item.
\subsubsection{Baselines}
To verify the efficacy of SMORE, we benchmark against several state-of-the-art (STOA) recommender models. These baselines fall into two main categories: General recommenders, which focus solely on user-item interaction data to provide recommendations, and multi-modal recommenders that leverage both historical data and the multi-modal features of each item.\\ 
\textbf{i) General Recommenders}: The following STOA models that include STOA matrix factorization (MF) model (BPR~\cite{rendle2012bpr}) and a graph-based model (LightGCN~\cite{he2020lightgcn}) are chosen for comparison.\\
\textbf{ii) Multi-modal Recommenders}: To ensure robust evaluation of the proposed model, several STOA MRSs have been selected for comparison, including the MF model~(VBPR\cite{he2016vbpr}) and graph-based models (MMGCN~\cite{wei2019mmgcn}, GRCN~\cite{wei2020graph}, SLMRec~\cite{tao2022self}, BM3~\cite{zhou2023bootstrap}, MGCN~\cite{yu2023multi}, FREEDOM~\cite{zhou2023tale}).
\subsubsection{Evaluation Standards.}
To ensure consistency, we adhere to existing works~\cite{zhang2021mining,zhou2023bootstrap} and divide the interaction data into 80\% for training, 10\% for validating, and 10\% for testing. Furthermore, we adopted the all-ranking protocol to evaluate top-K recommendation performance, using two widely-used metrics: Recall@K and NDCG@K. The results were reported for all users in the test set.
\subsubsection{Implementation Details.}
We utilized the unified open-source MMRec framework ~\cite{zhou2023mmrec} for developing the proposed model and replicating existing recommenders. For each of the selected baselines, the hyperparameters were tuned in line with the optimal configurations reported in the respective published papers. To further ensure impartiality, we complied with existing works~\cite{zhou2023bootstrap, zhou2023tale} and deployed the same seed across all baseline implementations and fixed the dimension of both the users and items at 64. We initialized all training parameters using the Xavier~\cite{glorot2010understanding} technique and adopted the Adam optimizer~\cite{kingma2014adam}. The training process employed a fixed batch size of 2,048 and was conducted over 1,000 epochs. Early stopping was activated after 20 consecutive steps without improvement on the validation set, with Recall@20 being the indicator metric.

\subsection{Effectiveness of SMORE (RQ1)}
With reference to Table~\ref{tab: Overall Performance Comparison}, comparisons with highly-competitive general and multi-modal recommenders reveal that:

\textbf{The proposed model consistently outperforms all baselines, including general and multi-modal recommenders.} We posit the improvement arises from SMORE's ability to capitalize on multi-modalities for inferring accurate uni-modal and fusion preferences. By leveraging the discriminative spectral characteristics and global perspective inherent in the frequency domain, universal patterns across different modalities are proficiently captured, while poor performance due to cross-modality noise is mitigated by the dynamic filter through effective attenuation and suppression. Furthermore, the multi-modal graph learning module empowers SMORE to encode high-order collaborative and semantically associated, preference-related modality features. To model the users' diverse modality preferences reliably, SMORE exploits the enriched universal fused signals to regulate the uni-modal features, ensuring that the uni-modal and fusion preferences are optimally balanced and accurately aligned with real-world scenarios.

\textbf{Graph-based recommenders that fuse modalities directly are evidently less effective.} In some instances, general models such as LightGCN achieve higher performance than MRSs such as MMGCN, which utilizes direct summation for fusion. This result suggests the existence of noise due to multi-modalities~(as illustrated in Fig.~\ref{Case Study}) and that direct fusion can adversely impact the performance of multi-modal recommenders.

\begin{table}[t]
\caption{Performance Comparison on multi-modalities}
    \centering
    \begin{tabular}{llcccc}
        \toprule
        Datasets & Modality & R@10 & R@20 & N@10 & N@20 \\
        \midrule
        \multirow{4}{*}{Baby} 
        & Text & 0.0646 & 0.0996 & 0.0341 & 0.0431 \\
        & Visual & 0.0533 & 0.0854 & 0.0290 & 0.0373 \\
        & Fusion & 0.0625 & 0.0964 & 0.0331 & 0.0418 \\
        & Full & \textbf{0.0680} & \textbf{0.1035} & \textbf{0.0365} & \textbf{0.0457} \\
        \midrule
        \multirow{4}{*}{Sports} 
        & Text & 0.0727 & 0.1099 & 0.0392 & 0.0488\\
        & Visual & 0.0592 & 0.0903 & 0.0323 & 0.0404 \\
        & Fusion & 0.0729 & 0.1091 & 0.0392 & 0.0486 \\
        & Full & \textbf{0.0762} & \textbf{0.1142} & \textbf{0.0408} & \textbf{0.0506}\\
        \midrule
        \multirow{4}{*}{Clothing} 
        & Text & 0.0631 & 0.0945 & 0.0343 & 0.0422 \\
        & Visual & 0.0443 & 0.0661 & 0.0241 & 0.0296 \\
        & Fusion & 0.0621 & 0.0937 & 0.0342 & 0.0422\\
        & Full & \textbf{0.0659} & \textbf{0.0987} & \textbf{0.0360} & \textbf{0.0443}\\
        \bottomrule
    \end{tabular}
    \label{tab:impact of modalities}
\end{table}
\begin{figure}[t]
    \centering
    \includegraphics[scale=0.260]{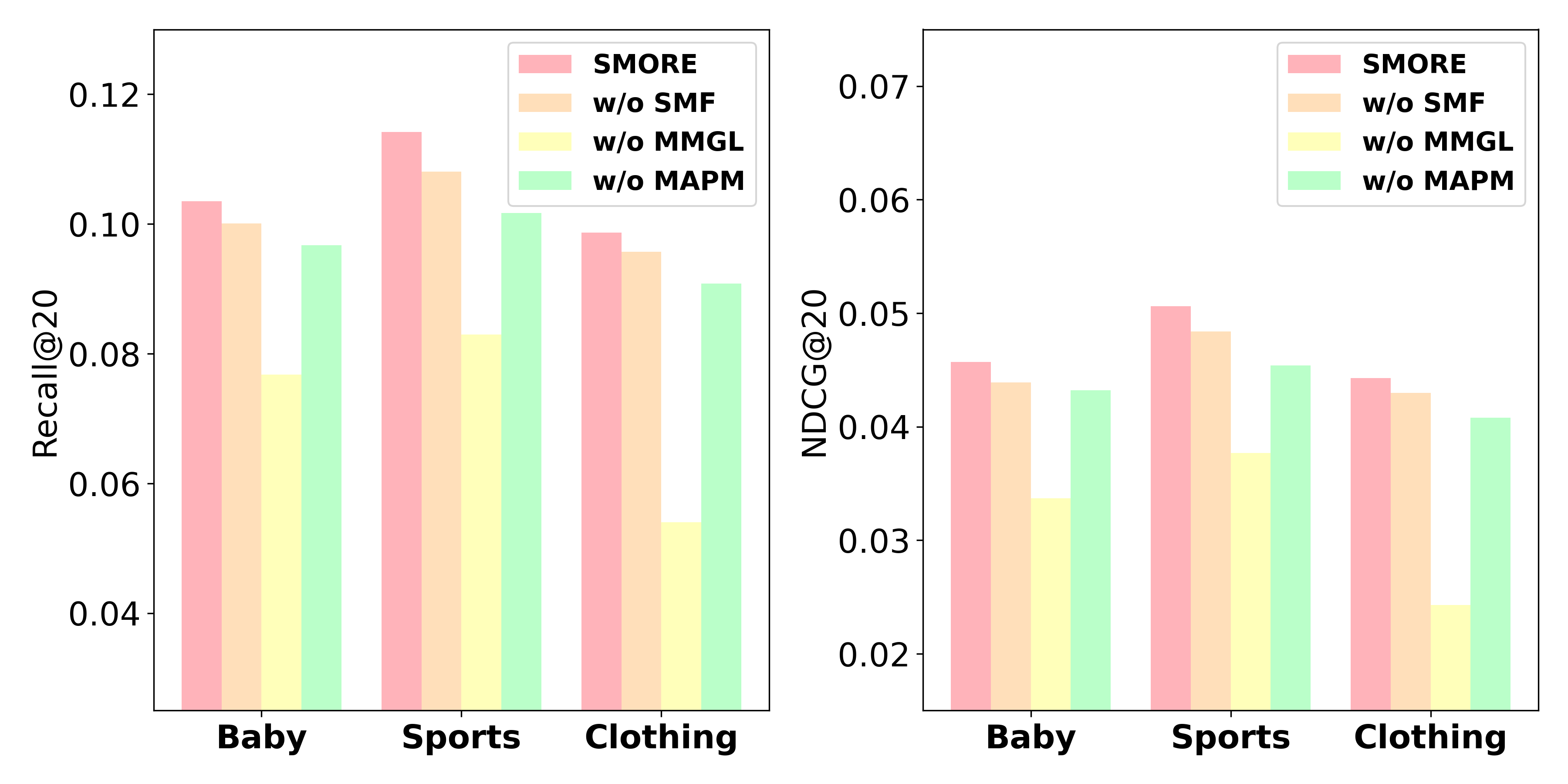}
    \caption{Ablation studies on the proposed SMORE}
    \label{SMORE Ablation}
\end{figure}
\textbf{To a certain extent, indirect integration of modality features may help to reduce modality noise.} Unlike VBPR, which directly injects modality features into ID representations, GRCN exhibits moderate performance improvement by relying on modality features implicitly to enhance the interaction graph. From an alternate perspective, the two MGCN and FREEDOM baselines exhibit enhanced performance by examining the latent structures of items through various modalities. We hypothesize that such a notable performance is attributed to MGCN's uni-modal noise reduction through behavioral injection and FREEDOM's degree-aware edge pruning strategy for denoising the interaction graph. Nonetheless, these models share a common vulnerability\textemdash they do not explore the complementary and universal features effectively across different modalities. In contrast, SMORE captures both uni-modal and fusion signals explicitly and can discern users' varying degrees of preferences accurately, resulting in its superior performance.

\subsection{Ablation Studies (RQ2)}
To ascertain the effectiveness of each component, we segment the proposed model into four distinct variants: (i) SMORE, (ii) SMORE without spectrum modality fusion, (iii) SMORE without multi-modal graph learning, and (iv) SMORE without modality aware preference module. Results presented in Fig.~\ref{SMORE Ablation} highlight that:

\textbf{Removing any key components leads to performance deterioration.} It is evident that every key component plays a significant role in SMORE, collectively contributing to its superior performance. For instance, spectrum modality fusion is responsible for fusing and mitigating cross-modality noise through attenuation and suppression, while multi-modal graph learning and modality-aware preference modules aim to encode semantically associative universal signals and decipher accurate user preference, respectively. Hence, omitting any of these modules degrades the performance of the proposed model.

\begin{figure}[t]
    \centering
    \includegraphics[scale=0.140]{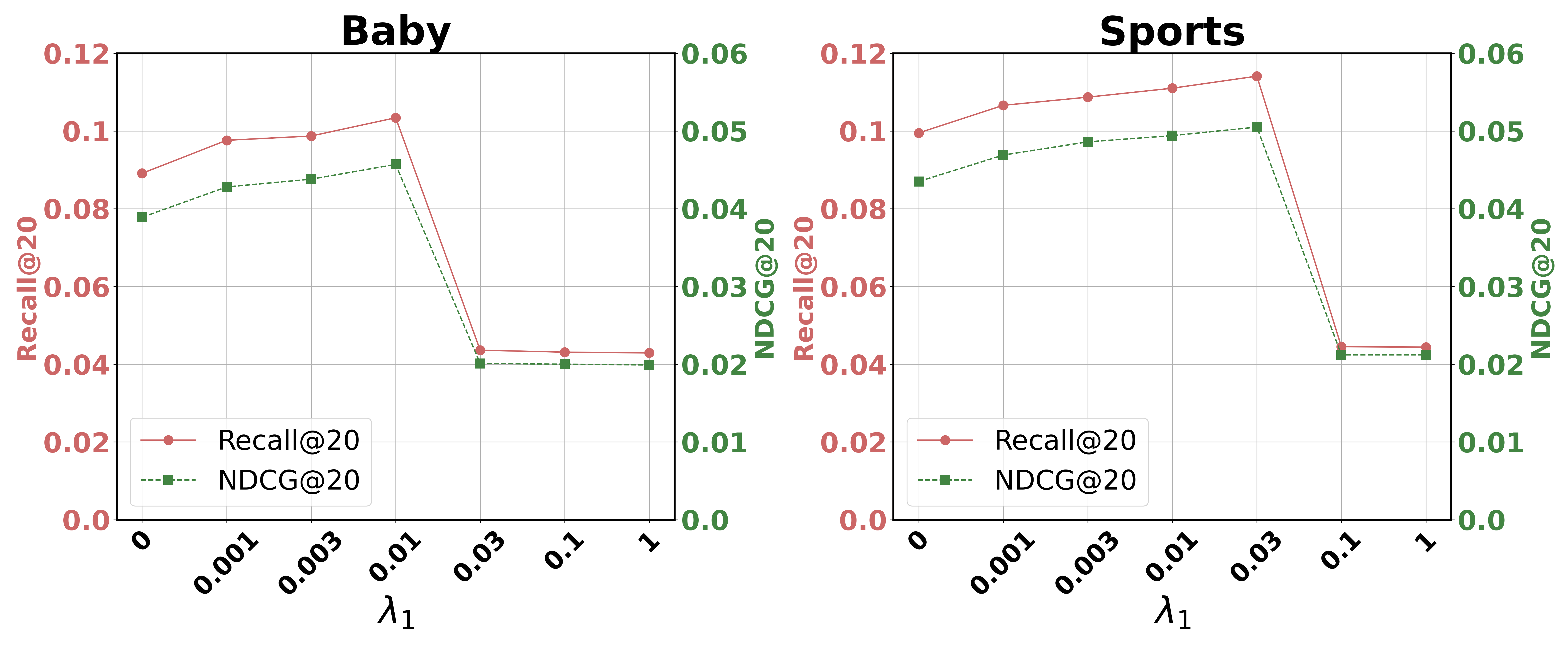}
    \captionsetup{skip=-2pt} 
    \caption{Variation of SMORE with $\lambda_1$}
    \label{CL_loss_plots}
\end{figure}
\begin{figure}[t]
    \centering
    \includegraphics[scale=0.160]{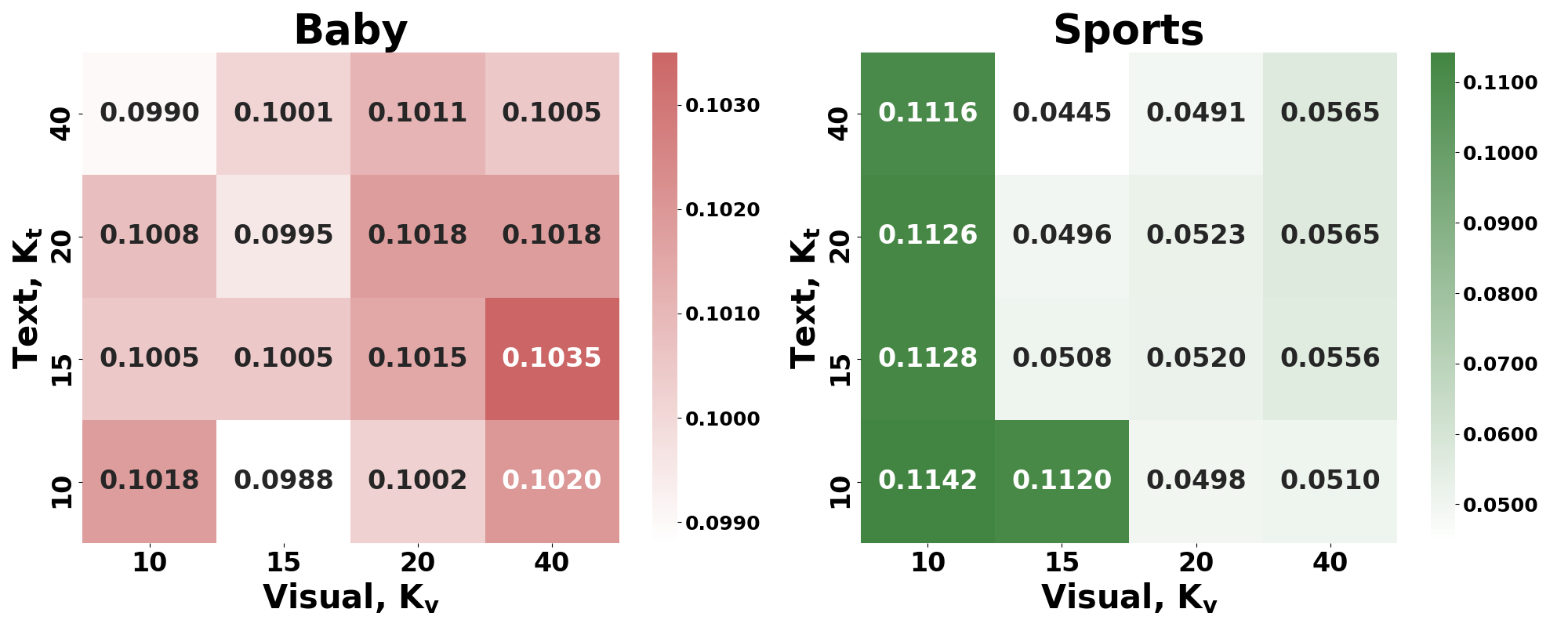}
    \caption{Variation of SMORE with $K_m$}     
    \label{KNN_K_plots}
\end{figure}
\textbf{MMGL plays a quintessential role in enhancing the overall performance.} We observe that, out of the four variants, the absence of MMGL results in a significant decline in performance across all datasets. This trend underscores the importance of encoding both high-order collaborative signals and the universal associative features derived from the denoised uni-modal and fused features, 
which collectively bolsters the overall performance of SMORE.

To further assess the impact of each modality, we perform experiments under a variety of input conditions: \textit{text} comprising textual (sequential) information, \textit{visual} consisting of pictorial (spatial) information, \textit{fusion} including only the fused (complementary) information, and \textit{full} encompassing both uni-modal and fusion information. Results tabulated in Table~\ref{tab:impact of modalities} indicate the following:

\textbf{The omission of any modalities results in reduced performance.} Excluding any modality reduces SMORE's capability to decipher users' diverse modality preferences. This observation provides unequivocal evidence that SMORE can leverage the uni-modal and complementary features encapsulated in the given modalities effectively while mitigating issues associated with noise contamination\textemdash a problem inherent in existing models~\cite{zhang2021mining,zhang2022latent}. This also reliably substantiates the necessity of modeling both uni-modal and fusion preferences, which serve as a realistic representation of real-world scenarios.

\textbf{Among the first three variants, SMORE demonstrates notable performance in utilizing sequential text information.} It has been shown in existing studies that the use of text usually leads to a significant degradation in terms of performance due to irrelevant information~\cite{yu2023multi}. On the contrary, we observe from Table~\ref{tab: Overall Performance Comparison} that utilizing text solely exhibits higher performance than STOA models (e.g., MGCN and FREEDOM), which utilize both modalities. This observation highlights the denoising capability of SMORE in capturing essential uni-modal features.

\begin{figure}[t]
    \centering
    \includegraphics[scale=0.27]
    {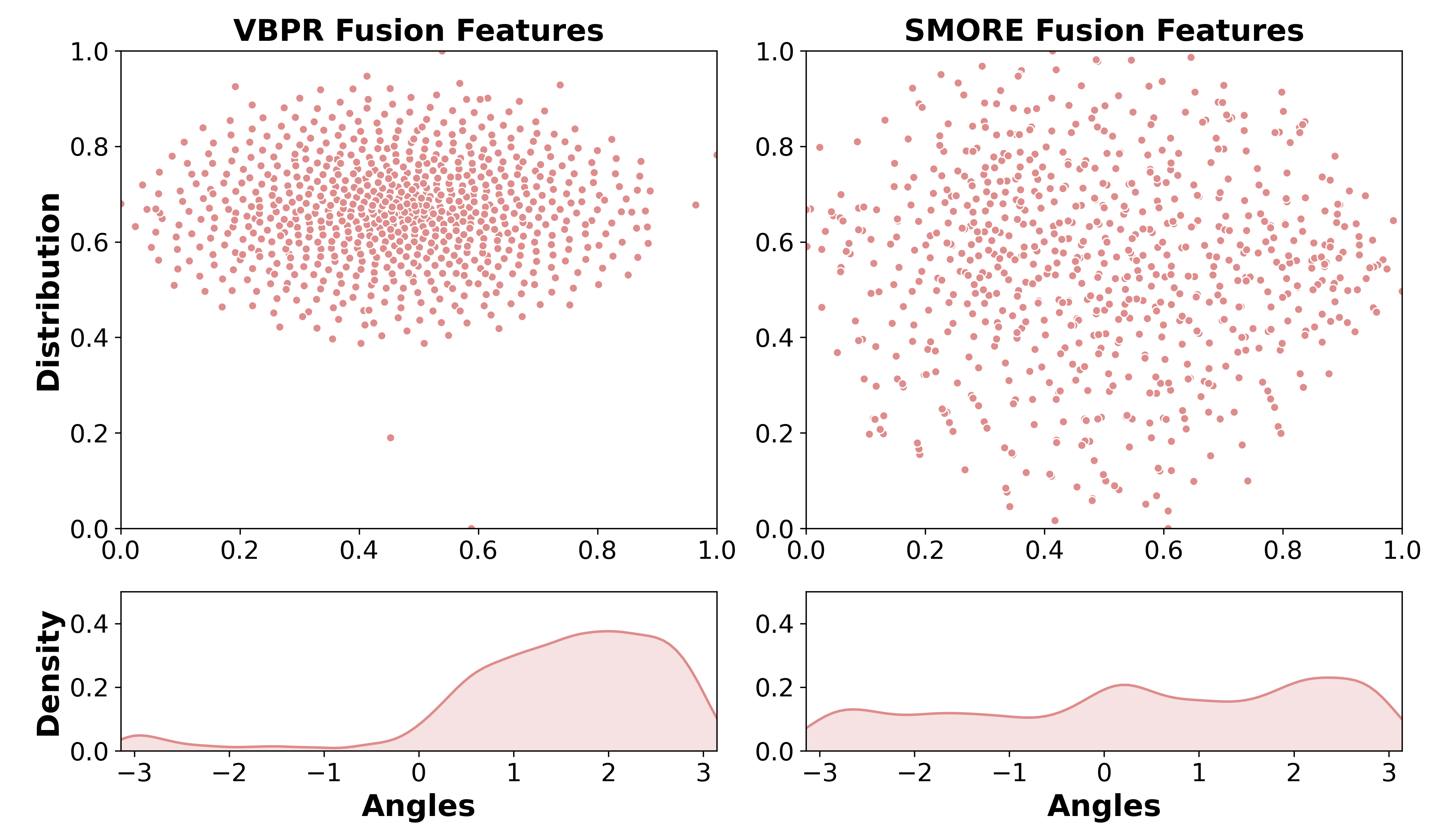}
    \caption{Distribution of fusion features for Baby Dataset}
    \label{Baby_TSNE}
    \vspace{-5mm}
\end{figure}
\subsection{Selection of Key Hyperparameters (RQ3)}
With reference to Fig.~\ref{CL_loss_plots}, we investigate the primary hyperparameter $\lambda_1$ of SMORE, which governs the influence of contrastive task by varying $0 \leq \lambda_1 \leq 1$. Results reveal that the omission of contrastive loss ($\lambda_1 = 0$) degrades the performance of the proposed model to a considerable degree, implying the beneficial impact of incorporating an auxiliary task for self-supervision and representation alignment. On the other hand, setting an excessively high value of $\lambda_1=1$ results in the most significant degradation due to the model placing undue emphasis on the auxiliary task. Notably, the best performance is achieved at $\lambda_1=0.01$ and $0.03$ for the baby and sports datasets, respectively. This observation implies that a small value is sufficient to enhance the recommendation task. 

We next assess the impact of $K_m$ defined in~(\ref{knn_sparsification}) on SMORE by varying $10 \leq K_m \leq 40$ for each modality. With reference to Fig.~\ref{KNN_K_plots}, findings from the sports dataset indicate that a small value ($K_m=10$) for both visual and text is sufficient to capture relevant uni-modal and fusion associative signals. However, as $K_v \geq 20$, performance degradation persists, while variations in $K_{t}$ generally do not compromise the performance. On the contrary, for the Baby dataset, we noted a minor variation in trend\textemdash while setting low value ($K_t = 15$) for text coheres with the prior finding, we observe that the optimal attained performance occurs when $K_v$ is set to a high value ($K_v = 40$). This suggests that a low value of $K$ may not always be effective across different datasets, and setting it too low may inadvertently discard essential neighbors.

\subsection{Impact of fusion in frequency domain (RQ4)}
\label{sec:fusion}
\begin{figure}[t]
    \centering
    \includegraphics[scale=0.27]
    {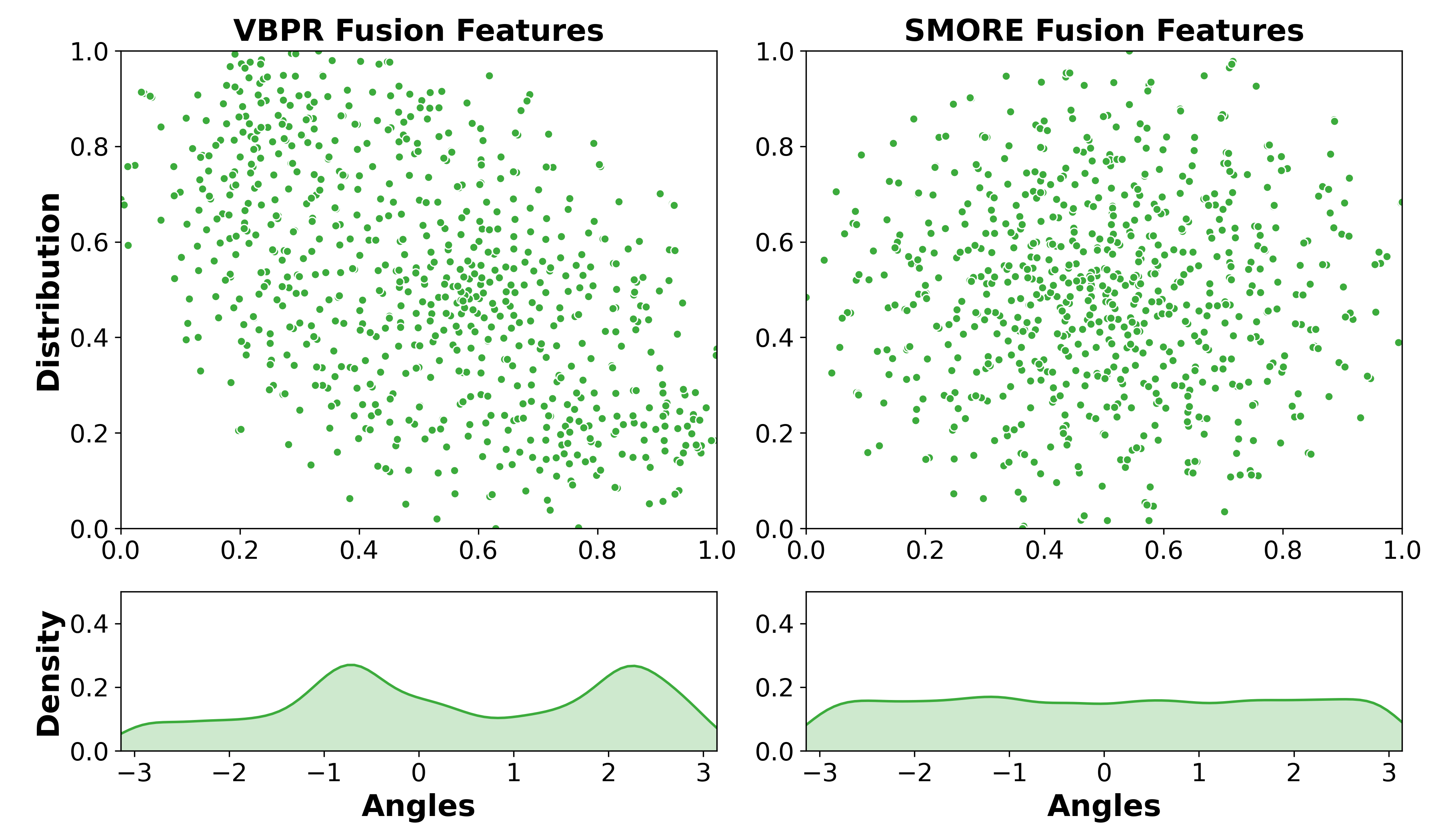}
    \caption{Distribution of fusion features for Sports Dataset}
    \label{Sports_TSNE}
    \vspace{-3.5mm}
\end{figure}

To verify the quality of the fusion features captured by the proposed model, we visualize the fusion features by first performing dimensionality reduction using t-SNE~\cite{van2008visualizing} to map the high-dimensional fusion features into 2-D, as illustrated in Figs.~\ref{Baby_TSNE} and~\ref{Sports_TSNE}. The embedding distributions are then plotted using Gaussian kernel density estimation, with the unit hypersphere $S^1$ showing the estimation of $\arctan(y, x)$ depicted at the bottom of each figure. For comparison, we used the classical VBPR fusion model as a benchmark. 

As depicted for the Baby dataset, we observe that the fusion features learned by SMORE result in a uniformly distributed structure implying that each item is characterized by its own distinct fusion semantic associations. On the other hand, we note that the distribution of VBPR is significantly condensed, highlighting the representation degeneration problem~\cite{gao2019representation, qiu2022contrastive}. This phenomenon is linked to the cross-modality noise amplification illustrated in Fig.~\ref{Case Study}, which severely corrupts and limits the expressivity of the representation after fusion. Similarly, Fig.~\ref{Sports_TSNE} exemplifies comparable patterns that mirror previous observations. While VBPR displays a lower level of degeneration issue, the proposed model attains greater uniformity, with distributions that are evenly spread. These results clearly substantiate the capability of SMORE in fusing different modalities and mitigating the cross-modality noise in the frequency domain, validating the efficacy of our fusion approach.

\section{Conclusion}
In this work, we aim to reduce modality noise by harnessing the discriminative spectrum property and global perspective inherent in the frequency domain for modality fusion and uni-modal denoising. Specifically, the proposed multi-modal SMORE recommender effectively captures both uni-modal and fusion preferences while actively suppressing modality noise. By leveraging the discriminative modality spectrum property, we proposed an effective approach to attenuate and suppress cross-modality noise during fusion. To explore the item latent structures, we introduced a new multi-modal graph learning module to distill long-range collaborative and semantic associated universal patterns among similar items. Finally, mirroring real-world scenarios, where users often display a mixture of multi-modal preferences, we designed a modality-aware preference module that effectively balances the uni-modal and fused representations, enabling a precise capture of users' uni-modal and fusion preferences.


\bibliographystyle{ACM-Reference-Format}
\bibliography{SMORE}

\appendix

\end{document}